\documentclass[prd,twocolumn,twoside,preprintnumbers,superscriptaddress,nofootinbib]{revtex4}

\usepackage{amsmath}
\usepackage{graphicx}
\usepackage{dcolumn}
\usepackage{xspace}
\usepackage{color}
\usepackage{url}
\usepackage[utf8]{inputenc}
\usepackage{epstopdf}

\newcommand{\bea}{\begin{eqnarray}}
\newcommand{\eea}{\end{eqnarray}}

\newcommand{\RDs}{$R_{D^{(*)}}$}

\begin{document}
%%%%%%%%%%%%%%%%%%%%%%%%%%%%%%%%%
\preprint{PSI-PR-17-19, LPT-Orsay-17-74}

\title{Searching for New Physics with \boldmath$b\to s\tau^+\tau^-$\unboldmath processes}

\author{Bernat Capdevila}
\email{bcapdevila@ifae.es}
\affiliation{Universitat Aut\`onoma de Barcelona, 08193 Bellaterra, Barcelona,\\
Institut de Fisica d'Altes Energies (IFAE), The Barcelona Institute of Science and Technology, Campus UAB, 08193 Bellaterra (Barcelona)}

\author{Andreas Crivellin}
\email{andreas.crivellin@cern.ch}
\affiliation{Paul Scherrer Institut, CH--5232 Villigen PSI, Switzerland}

\author{S\'ebastien Descotes-Genon}
\email{sebastien.descotes-genon@th.u-psud.fr}
\affiliation{Laboratoire de Physique Th\'eorique (UMR8627),\\
CNRS, Univ. Paris-Sud, Universit\'e Paris-Saclay, 91405 Orsay, France}

\author{Lars Hofer}
\email{hofer@fqa.ub.edu}
\affiliation{Departament de F\'{\i}sica Qu\`antica i Astrof\'{\i}sica (FQA),
Institut de Ci\`encies del Cosmos (ICCUB), Universitat de Barcelona (UB), Spain}

\author{Joaquim Matias}
\email{matias@ifae.es}
\affiliation{Universitat Aut\`onoma de Barcelona, 08193 Bellaterra, Barcelona,\\
Institut de Fisica d'Altes Energies (IFAE), The Barcelona Institute of Science and Technology, Campus UAB, 08193 Bellaterra (Barcelona)}

%%%%%%%%%%%%%%%%%%%%%%%%%%%%%%%%%%
%%%%%%%%%%%%%%%%%%%%%%%%%%%%%%%%%%
\begin{abstract}
In recent years, intriguing hints for the violation of Lepton Flavour 
Universality (LFU) have been accumulated 
in semileptonic $B$ decays, both in the neutral-current transitions $b\to s\ell^+\ell^-$ (i.e., $R_K$ and $R_{K^*}$) and the charged-current transitions $b\to c\ell^-\bar\nu_\ell$ (i.e., $R_D$, $R_{D^*}$ and $R_{J/\psi}$). LHCb has reported deviations from the Standard Model (SM) expectations in $b\to s\mu^+\mu^-$ processes as well as in the ratios $R_K$ and $R_{K^*}$, which together point at New Physics (NP) affecting muons with a high significance. Furthermore, hints for LFU violation in $R_{D^{(*)}}$ and $R_{J/\psi}$ point at large deviations from the SM in processes involving tau leptons. Together, these hints for NP motivate the possibility of huge LFU-violating effects in $b\to s\tau^+\tau^-$ transitions. In this article we predict the branching ratios of $B\to K\tau^+\tau^-$, $B\to K^{*}\tau^+\tau^-$ and $B_s\to \phi \tau^+\tau^-$ taking into account NP effects in the Wilson coefficients $C_{9(')}^{\tau\tau}$ and $C_{10(')}^{\tau\tau}$. Assuming a common NP explanation of $R_{D^{}}$ , $R_{D^{(*)}}$ and $R_{J/\psi}$, we show that  a very large enhancement of $b\to s\tau^+\tau^-$ processes, of around three orders of magnitude compared to the SM, can be expected under fairly general assumptions. We find that the branching ratios of $B_s\to \tau^+\tau^-$, $B_s\to \phi \tau^+\tau^-$ and $B\to K^{(*)}\tau^+\tau^-$ under these assumptions are  in the observable range for LHCb and Belle II.
\end{abstract}
\pacs{}

%%%%%%%%%%%%%%%%%%%%%%%%%%%%%%%%%
%%%%%%%%%%%%%%%%%%%%%%%%%%%%%%%%%
\maketitle

%%%%%%%%%%%
\section{Introduction}
\label{intro}
%%%%%%%%%%%

After the discovery of the Higgs boson, the search for physics beyond the Standard Model has become a particularly relevant subject. While the LHC has not observed any new fundamental particle beyond the SM ones directly so far, several intriguing hints of New Physics (NP) in semileptonic $B$ decays arose recently. 

On the one hand, the $b\to s\mu^+\mu^-$ flavour-changing neutral current is loop suppressed in the SM and it has been measured by several experiments, showing a collection of deviations from the SM in angular observables and branching ratios. Moreover, the comparison of $b\to s\mu^+\mu^-$ and $b\to se^+e^-$ through the measurements of $R_K$~\cite{Aaij:2014ora} and $R_{K^*}$~\cite{Aaij:2017vbb} for several values of the dilepton invariant mass suggest a significant violation of Lepton Flavour Universality (LFU). All these deviations can be explained in a model-independent approach by NP contributions to Wilson coefficients associated with the operators describing $b\to s\mu^+\mu^-$ transitions, providing a consistent description of the observed pattern. A recent combined analysis of these observables~\cite{Capdevila:2017bsm} indeed singles out some NP scenarios preferred over the SM with a significance at the $5\,\sigma$ level.This confirms the scenarios already highlighted in earlier analyses, mainly restricted to $b\to s\mu^+\mu^-$ processes~\cite{Descotes-Genon:2013wba,Descotes-Genon:2015uva,Altmannshofer:2015sma,Hurth:2016fbr}. The significance for these NP scenarios considering only the LFU-violating observables $R_K$ and $R_{K^*}$  (and excluding other $b\to s\mu^+\mu^-$ processes) is at the 3-$4\,\sigma$ level~\cite{Altmannshofer:2017yso,DAmico:2017mtc,Geng:2017svp,Ciuchini:2017mik,Hiller:2017bzc}. The corresponding violation of LFU between muons and electrons is indeed significant, around $25\%$ at the level of some of the Wilson coefficients.

On the other hand, measurements of the $b\to c\ell^-\bar\nu_\ell$ charged current have also shown interesting patterns of deviations, even though these are tree-level processes in the SM which are in general less sensitive to NP. The ratios $R_{D^{(*)}}$, which measure LFU violation in the charged current by comparing the tau mode to light lepton ($e,\mu$) modes, differ from their SM predictions by a combined significance of approximately $4\,\sigma$~\cite{Amhis:2016xyh}. The effect related to tau leptons in $R_{D^{(*)}}$ corresponds to an $O(10\%)$ effect at the amplitude level, assuming its interference with the SM. Recently, LHCb released results for the ratio $R_{J/\psi}$~\cite{Aaij:2017tyk} which measures LFU violation in $b\to c\ell^-\bar\nu_\ell$ as well. Again, even though the error is large, the experimental central value  exceeds the SM prediction  in agreement with the expectations from $R_{D^{(*)}}$~\cite{Watanabe:2017mip,Chauhan:2017uil,Dutta:2017wpq,Alok:2017qsi}.

Taking into account these hints for NP, we might expect to have a large LFU violation in the neutral current involving tau leptons, i.e., in $b\to s\tau^+\tau^-$ transitions. In fact, it has been shown in Refs.~\cite{Alonso:2015sja,Crivellin:2017zlb,Calibbi:2017qbu} that one can expect an enhancement of up to three orders of magnitude compared to the SM predictions in $b\to s\tau^+\tau^-$ processes if one aims at explaining the central value of $R_{D^{(*)}}$. So far, among the possible processes, only LHCb searched for $B_s\to\tau^+\tau^-$~\cite{Aaij:2017xqt}
\begin{equation}
{\rm{Br}}{\left( {B_s \to {\tau ^ + }{\tau ^ - }} \right)_{{\rm{EXP}}}} \le 6.8 \times {10^{ - 3}}\,,
\end{equation}
and BaBar performed an analysis of $B \to {K}{\tau^+\tau^-}$~\cite{TheBaBar:2016xwe}
\begin{equation}
{\rm{Br}}{\left( {B \to {K}{\tau^+\tau^-}} \right)_{{\rm{EXP}}}} \le 2.25 \times {10^{ - 3}}\,.
\end{equation}
A search for $B \to {K^{(*)}}{\tau^+\tau^-}$ or $B_s \to {\phi}{\tau^+\tau^-}$ should be possible at LHCb: compared to the case of $B_s \to {\tau ^ + }{\tau ^ - }$, these analyses involve more tracks (originating from the $K$, $K^*$ or $\phi$ mesons) that can be reconstructed. In addition, the Belle experiment has not analysed their data for $b\to s\tau^+\tau^-$ transitions yet and the upcoming Belle II experiment should be able to improve significantly on the measurement of $B \to {K^{(*)}}{\tau^+\tau^-}$ decays: an $e^+e^-$ experiment such as Belle II can be expected to be more efficient in reconstructing $B$ decays to tau leptons than LHCb. Since Belle II is expected to run at the $\Upsilon(4S)$ resonance, it will not study $B_s \to {\tau ^ + }{\tau ^ - }$ whereas $B \to {K^{(*)}}{\tau^+\tau^-}$ are golden modes for finding NP at this facility. There are thus good experimental prospects for these transitions in the coming years.

On the theory side, $b\to s\tau^+\tau^-$ processes have received a limited attention so far. Within the SM, the $B_s\to\tau^+\tau^-$ branching ratio is known very precisely~\cite{Bobeth:2013uxa,Bobeth:2014tza}
\begin{equation}
	{\rm{Br}}{\left( {B_s \to {\tau ^ + }{\tau ^ - }} \right)_{{\rm{SM}}}} = \left( {7.73 \pm 0.49} \right) \times {10^{ - 7}}\,, \label{brbstautauSM}
\end{equation}
whereas the $b\to s\tau^+\tau^-$ processes $B\to K^*\tau^+\tau^-$, $B\to K\tau^+\tau^-$ and $B_s\to \phi\tau^+\tau^-$ have not been considered in detail until recently, especially concerning the impact of NP contributions.
Only the branching ratio for $B\to K\tau^+\tau^-$ was estimated in Ref.~\cite{Bobeth:2011st} including NP effects. Recently, an analysis of branching ratios and tau polarisations in $b\to s\tau^+\tau^-$ was performed to determine the sensitivity to NP contributions to the  Wilson coefficients~\cite{Kamenik:2017ghi}. 
 
Within the SM, the branching ratios for $B\to K^*\tau^+\tau^-$ and $B_s\to \phi\tau^+\tau^-$ are known to be of $O(10^{-7})$~\cite{Hewett:1995dk,Bouchard:2013mia,Kamenik:2017ghi} and the inclusive $B\to X_s\tau^+\tau^-$ process was assessed in Refs.~\cite{Guetta:1997fw,Bobeth:2011st}. Ref.~\cite{Bobeth:2011st} also studied the indirect constraints on $b\to s\tau^+\tau^-$ operators, finding that the constraints on NP contributions are very loose once the effects in $b \to s\tau^+\tau^-$ and $b \to d\tau^+\tau^-$ transitions are correlated such that the stringent bounds from $\Delta \Gamma_s/\Delta \Gamma_d$ are avoided. Interestingly, sizable effects in analogous $b\to d\tau^+\tau^-$ operators~\cite{Bobeth:2014rda} could help solving the long-standing anomaly in the like-sign dimuon asymmetry measured by the D{\O} experiment~\cite{Lenz:2010gu,Lenz:2012az}.

In this article we look in detail at the $b\to s\tau^+\tau^-$ processes $B_s\to\tau^+\tau^-$, $B\to K^*\tau^+\tau^-$, $B\to K\tau^+\tau^-$ and $B_s\to \phi\tau^+\tau^-$. We will express their branching ratios in terms of the Wilson coefficients $C_{9(')}$ and $C_{10(')}$.
 In order to compute these processes we will use the same approach as in Ref.~\cite{
 Descotes-Genon:2015uva} to compute $b \to s \mu\mu$ observables, substituting muons by taus and taking the relevant form factors  in the $q^2$-region for the $\tau^+\tau^-$ invariant mass where these decays are allowed kinematically. Since the mass of the tau leptons cannot be neglected compared to the $B$ meson, this region is much smaller than for decays to light leptons and we will consider the branching ratios only in the equivalent of the high-$q^2$ region (or low recoil) for lighter leptons.

The article is structured as follows: In Sec.~\ref{Sec:EFT}, we consider the generic effects of NP originating from vector operators. In Sec.~\ref{Sec:RD} we correlate the effects in $b\to s\tau^+\tau^-$ to $R_D$ and $R_{D^*}$ and study the impact on branching ratios, before we conclude in Sec.~\ref{Sec:conclusion}.

\section{EFT approach to $b\to s\tau^+\tau^-$}\label{Sec:EFT}

In this section we express the branching ratios for our $b\to s\tau^+\tau^-$ processes as functions of $C_{9(')}^{\tau\tau}$ and $C_{10(')}^{\tau\tau}$ and calculate the SM predictions. We define our effective Hamiltonian in the following way, focusing on the relevant operators for our discussion
\begin{eqnarray}
H_{\rm eff}(b\to s\tau\tau)&=&- \dfrac{ 4 G_F }{\sqrt 2}V_{tb}V_{ts}^{*} \sum\limits_{a} C_a O_a\,,\\
 {O_{9(10)}^{\tau\tau}} &=&\dfrac{\alpha }{4\pi}[\bar s{\gamma ^\mu } P_L b]\,[\bar\tau{\gamma _\mu }(\gamma^5)\tau] \,,\\
 {O_{9'(10')}^{\tau\tau}} &=&\dfrac{\alpha }{4\pi}[\bar s{\gamma ^\mu } P_R b]\,[\bar\tau{\gamma _\mu }(\gamma^5)\tau] \,,
\label{eq:effHam}
\end{eqnarray}
where $C_{9}^{\rm SM} \approx  4.1$ and $C_{10}^{\rm SM} \approx  - 4.3$ at  the scale $\mu=4.8$ GeV~\cite{Bobeth:1999mk,Huber:2005ig,DescotesGenon:2011yn}, $P_{L,R}=(1\mp \gamma_5)/2$, and the chirality-flipped coefficients have negligible contributions in the SM.

Besides ${{\rm{Br}}\left( {{B_s} \to \tau^+ \tau^- }\right)}_{\rm SM}$ given in Eq.~(\ref{brbstautauSM}) we
use the approach and inputs in Refs.~\cite{Descotes-Genon:2013vna,Descotes-Genon:2014uoa,Descotes-Genon:2015uva,Capdevila:2017bsm} to compute the other processes of interest. Averaging over the charged and the neutral modes for ${B} \to K^{(*)} \tau^+ \tau^-$ we find
\begin{eqnarray}\label{eq:SMval1}
{\rm{Br}}\left( {{B} \to K \tau^+ \tau^- } \right)_{\rm SM}^{[15,22]}&=& \left(1.20\pm 0.12\right) \times 10^{-7}\,,\\
{\rm{Br}}\left( {{B} \to K^* \tau^+ \tau^- } \right)_{\rm SM}^{[15,19]}&=& \left(0.98\pm 0.10\right) \times 10^{-7}\,,\\
{\rm{Br}}\left( {{B_s} \to \phi \tau^+ \tau^- } \right)_{\rm SM}^{[15,18.8]}&=& \left(0.86\pm 0.06\right) \times 10^{-7}\label{eq:SMval3}
\end{eqnarray}
The superscript denotes the $q^2$-range for the dilepton invariant mass. This broad bin is chosen to leave out the $\psi(2S)$ resonance allowing the use of quark-hadron duality. As discussed in our previous works, our error budget includes in particular a conservative estimate of 10\% for duality violation effects, while estimates based on resonance models~\cite{Beylich:2011aq} yield violations around 2\%. 

In order to assess the structure of the branching ratios including beyond the SM effects, we parametrize both the central value and uncertainty of the branching ratio in each channel as quadratic polynomials in $C_9^\text{NP}$, $C_{10}^\text{NP}$, $C_{9'}$ and $C_{10'}$. The values of the polynomial coefficients are estimated by performing a fit to our theoretical predictions computed on an evenly spaced grid in the parameter space $\left\{C_9^\text{NP},C_{10}^\text{NP},C_{9'},C_{10'}\right\}$, with 300 points each in the ranges [-2,2], [-2,2], [-1,1] and [-0.2,0.2], respectively.

\begin{widetext}
\begin{eqnarray}\label{eq:NPdep1}
10^7\times{\rm{Br}}\left({{B}\to K\tau^+\tau^-}\right)^{[15,22]}&=&\big(1.20+0.15\,C_9^\text{NP}-0.42\,C_{10}^\text{NP}+0.15\,C_9^\prime-0.42\,C_{10}^\prime+0.04\,C_9^\text{NP}C_9^\prime\nonumber\\ 
&&\qquad +0.10\,C_{10}^\text{NP}C_{10}^\prime+0.02\,C_9^{\text{NP}\,2}+0.05\,C_{10}^{\text{NP}\,2}+0.02\,C_9^{\prime\,2}+0.05\,C_{10}^{\prime\,2}\big)\nonumber\\
&& \pm
\big(0.12+0.02\,C_9^\text{NP}-0.04\,C_{10}^\text{NP}+0.01\,C_9^\prime-0.04\,C_{10}^\prime\nonumber\\ 
&&\qquad +0.01\,C_{10}^\text{NP}C_{10}^\prime+0.01\,C_{10}^{\text{NP}\,2}+0.08\,C_{10}^{\prime\,2}\big)\,,\\
10^7\times{\rm{Br}}\left({{B}\to K^*\tau^+\tau^-}\right)^{[15,19]}&=&\big(0.98+0.38\,C_9^\text{NP}-0.14\,C_{10}^\text{NP}-0.30\,C_9^\prime+0.12\,C_{10}^\prime-0.08\,C_9^\text{NP}C_9^\prime\nonumber\\ 
&&\qquad -0.03\,C_{10}^\text{NP}C_{10}^\prime+0.05\,C_9^{\text{NP}\,2}+0.02\,C_{10}^{\text{NP}\,2}+0.05\,C_9^{\prime\,2}+0.02\,C_{10}^{\prime\,2}\big)\nonumber\\
&& \pm \big(0.09+0.03\,C_9^\text{NP}-0.01\,C_{10}^\text{NP}-0.03\,C_9^\prime-0.01\,C_9^\text{NP}C_9^\prime\nonumber\\ 
&&\qquad -0.01\,C_9^\prime C_{10}^\prime+0.01\,C_9^{\prime\,2}-0.01\,C_{10}^{\prime\,2}\big)\,,\\
10^7\times{\rm{Br}}\left({{B_s}\to \phi\tau^+\tau^-}\right)^{[15,18.8]}&=&\big(0.86+0.34\,C_9^\text{NP}-0.11\,C_{10}^\text{NP}-0.28\,C_9^\prime+0.10\,C_{10}^\prime-0.08\,C_9^\text{NP}C_9^\prime\nonumber\\ 
&&\qquad -0.02\,C_{10}^\text{NP}C_{10}^\prime+0.05\,C_9^{\text{NP}\,2}+0.01\,C_{10}^{\text{NP}\,2}+0.05\,C_9^{\prime\,2}+0.01\,C_{10}^{\prime\,2}\big)\nonumber\\
&& \pm \big(0.06+0.02\,C_9^\text{NP}-0.02\,C_9^\prime+0.02\,C_{10}^{\prime\,2}\big)
\label{eq:NPdep3}
\end{eqnarray}
\end{widetext}

As expected, there is a limited dependence of the uncertainties on the values of the Wilson coefficients. In order to shorten the equations, we dropped the superscript $\tau\tau$ in the Wilson coefficients here.
Comparing our results with Ref.~\cite{Kamenik:2017ghi}, we find slightly lower central values for the SM (Eqs.~(\ref{eq:SMval1})-(\ref{eq:SMval3})). On the other hand, we obtain 
the same dependence of the central values on the NP contributions to the Wilson coefficients (Eqs.~(\ref{eq:NPdep1})-(\ref{eq:NPdep3})). 

As stated above, in this analysis we neglect the effects of scalar and tensor operators. This is justified since the current global analyses of $b\to s\ell^+\ell^-$ anomalies do not favour such contributions~\cite{Descotes-Genon:2013wba,Descotes-Genon:2015uva,Altmannshofer:2015sma,Hurth:2016fbr}. Moreover, the indirect bounds on the Wilson coefficients of scalar operators from ${{B_s} \to \tau^+ \tau^- }$ are much stronger than for $C_{9(')}$ and $C_{10(')}$~\cite{Bobeth:2011st} and therefore they cannot lead to comparably large and observable effects in ${B} \to K^{(*)} \tau^+ \tau^-$ or ${B_s} \to \phi \tau^+ \tau^-$. We also neglect tensor operators since they are not generated at the dimension-6 level for $b\to s\ell^+\ell^-$\cite{Alonso:2014csa,Aebischer:2015fzz}.

\begin{boldmath}
\section{Correlation with $R_{D^{(*)}}$ and $R_{J/\psi}$}\label{Sec:RD}
\end{boldmath}

It is interesting to correlate these results with the tree-level $b\to c\tau^-\bar\nu_\tau$ transition.
A solution of the $\sim 4\sigma$ anomaly in {\RDs} and $R_{J/\psi}$ requires a NP contribution of $\mathcal{O}(20\%)$ to the branching ratio of $B\to D^{(*)}\tau^-\bar\nu_\tau$, which is rather large given that these decays are mediated in the SM already at tree level. In order to comply with the $B_c$ lifetime~\cite{Alonso:2016oyd} and the $q^2$ distribution of \RDs~\cite{Freytsis:2015qca,Celis:2016azn,Ivanov:2017mrj}, a contribution to the SM operator $[\bar c\gamma^\mu P_L b][ \bar \tau \gamma_\mu P_L \nu_{\tau}]$ is favoured such that there is interference with the SM.
In principle, these constraints can be avoided with right-handed couplings~\cite{Li:2016vvp} (including possibly right-handed neutrinos~\cite{Becirevic:2016yqi}). However, no interference with the SM appears for such solutions, which require very large couplings close to the perturbativity limit, and we will not consider such solutions any further.

Since a NP contribution to the Wilson coefficient of the SM $V-A$ operator amounts only to changing the normalisation of the Fermi constant for $b\to s\tau^+\tau^-$ transitions, one predicts in this case:
\begin{equation}\label{eq:ratiosR}
R_{J/\psi}/R_{J/\psi}^{{\rm SM}}=R_D/R_D^{{\rm SM}}=R_{D^*}/R_{D^*}^{\rm SM}\,,
\end{equation}
which agrees well with the current measurements.

If NP generates this contribution from a scale much larger than the electroweak symmetry breaking scale~\cite{Buchmuller:1985jz,Grzadkowski:2010es}, the semileptonic decays involving only left-handed quarks and leptons are described by the two $SU(2)_L$-invariant operators
\begin{eqnarray}\label{eq:o1o3}
   \mathcal{O}^{(1)}_{ijkl}&=&[\bar Q_i\gamma_\mu Q_j][\bar L_k\gamma^\mu L_l],\nonumber\\ 
   \mathcal{O}^{(3)}_{ijkl}&=&[\bar Q_i\gamma_\mu \sigma^I Q_j][\bar L_k\gamma^\mu \sigma^I L_l],
\end{eqnarray}
where the Pauli matrices $\sigma^I$ act on the weak-isospin components of the quark (lepton) doublets $Q$ $(L)$. Note that there are no further dimension-six operators involving only left-handed fields and that dimension-eight operators can be neglected for NP around the TeV scale. This approach has been used to correlate Wilson coefficients of the effective Hamitlonian for both charged- and neutral-current transitions in various broad classes of NP models (some examples are found in Refs.~\cite{Alonso:2015sja,Calibbi:2015kma,Celis:2017doq,Buttazzo:2017ixm}). 

After electroweak symmetry breaking, these operators contribute to semileptonic $b\to c(s)$ decays involving charged tau leptons and tau neutrinos. Working in the down basis when writing the $SU(2)$ components of the operators in Eq.~(\ref{eq:o1o3}) (i.e., in the field basis with diagonal down quark mass matrices) we obtain
\begin{widetext}
\begin{eqnarray}
   C^{(1)}\mathcal{O}^{(1)}&\to&C_{23}^{(1)}\left([\bar{s}_L\gamma_\mu b_L][\bar{\tau}_L\gamma^\mu\tau_L]+
           [\bar{s}_L\gamma_\mu b_L][\bar{\nu}_\tau\gamma^\mu\nu_\tau]\right),\\
   C^{(3)}\mathcal{O}^{(3)}&\to&C_{23}^{(3)}\left(2V_{cs}[\bar{c}_L\gamma_\mu b_L][\bar{\tau}_L\gamma^\mu\nu_\tau]+[\bar{s}_L\gamma_\mu b_L][\bar{\tau}_L\gamma^\mu\tau_L]-
           [\bar{s}_L\gamma_\mu b_L][\bar{\nu}_\tau\gamma^\mu\nu_\tau]\right)+C_{33}^{(3)}\left(2V_{cb}[\bar{c}_L\gamma_\mu b_L][\bar{\tau}_L\gamma^\mu\nu_\tau]\right).\nonumber
\end{eqnarray}
\end{widetext}
where $C_{ij}^{(n)}$ denote the Wilson coefficients for $\mathcal{O}^{(n)}_{ij33}$. 

We neglect the effect of $C_{13}^{(3)}$ which would enter $b\to c\tau^-\bar\nu_\tau$ processes with a factor proportional to $V_{cd}$. But it would contribute even more dominantly to $b\to d\tau^+\tau^+$ and $b\to u\tau^-\bar\nu_\tau$ processes such as $B^-\to \tau^-\bar\nu_\tau$, where no deviation from the SM is observed~\cite{Charles:2015gya,Koppenburg:2017mad}. We will thus not consider this contribution any more.

As a consequence, we see that $b\to c\tau^-\bar\nu_\tau$ processes receive a NP contribution from $C_{33}^{(3)}$ also in scenarios with a flavour-diagonal alignment to the third generation, which would avoid any effects in down-quark FCNCs. However, due to the CKM suppression of this contribution, a solution of the $R_{D^{(*)}}$ anomaly
via this contribution requires a rather large $C_{33}^{(3)}$ coming into conflict with bounds from electroweak precision data~\cite{Feruglio:2016gvd} and direct LHC searches for $\tau^+\tau^-$ final states~\cite{Faroughy:2016osc}. 

\begin{figure*}[t]
\begin{center}
\begin{tabular}{cp{7mm}c}
\includegraphics[width=0.7\textwidth]{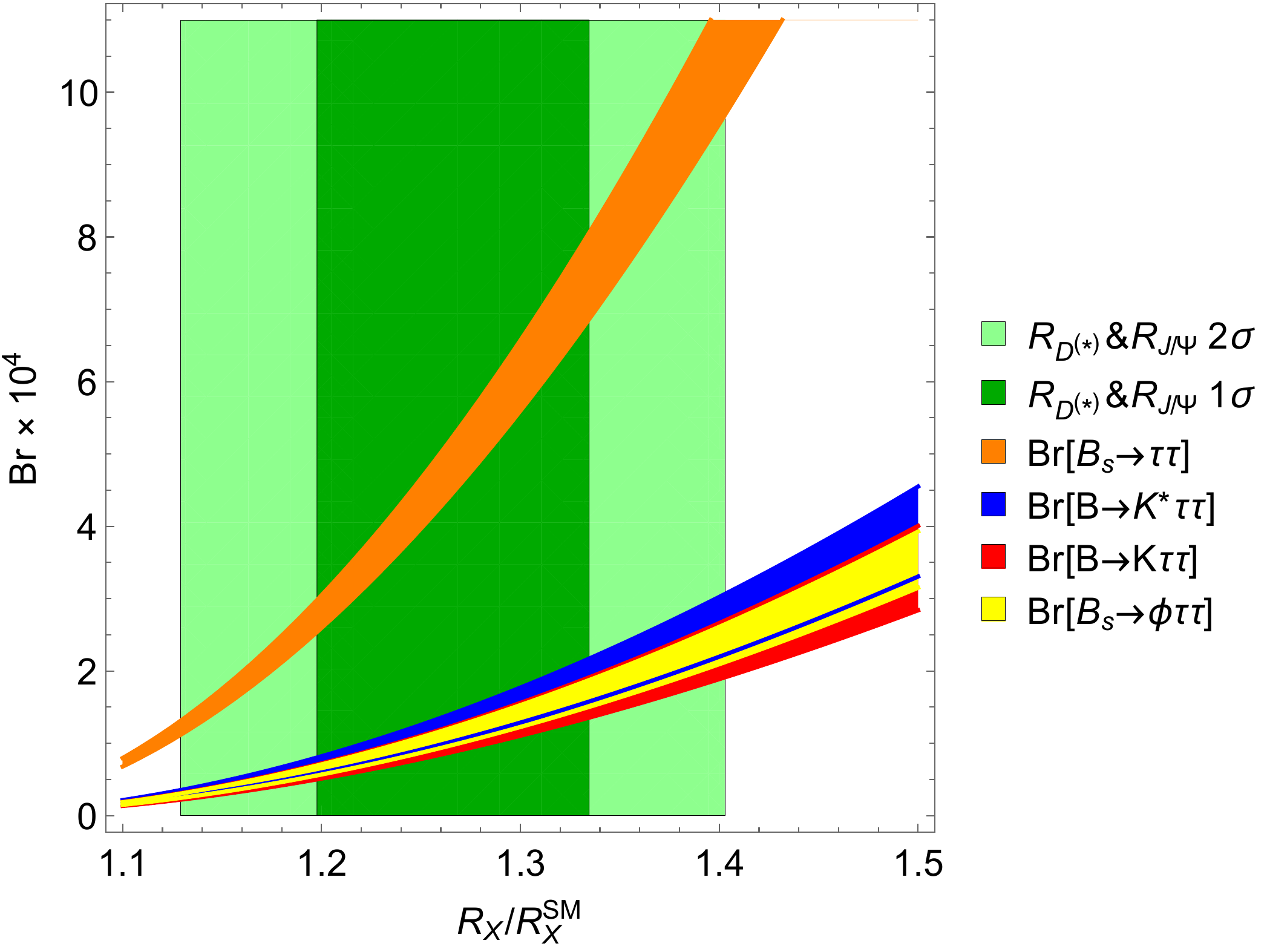}
\end{tabular}
\end{center}
\caption{Predictions of the branching ratios of the $b\to s\tau^+\tau^-$ processes (including uncertainties) as a function of $R_X/R_X^{\rm SM}$.}         
\label{fig:predsbstautau}
\end{figure*}

The $R_{D^{(*)}}$ anomaly can thus only be solved via $C_{23}^{(1,3)}$ which then must generate huge contributions to $b\to s\tau^+\tau^-$ and/or $b\to s\nu_\tau\bar{\nu}_\tau$ processes. 
The severe bounds on NP from $B\to K^{(*)}\nu\bar{\nu}$ (e.g., Ref.~\cite{Buras:2014fpa}) rule out large effects in $b\to s\nu\bar{\nu}$ and they can only be accommodated if the contribution from $C^{(3)}_{23}$ is approximately cancelled by the one from $C^{(1)}_{23}$, implying $C^{(1)}_{23}\approx C^{(3)}_{23}$~\cite{Buttazzo:2017ixm}. Such a situation can for instance be realized by a vector leptoquark singlet~\cite{Alonso:2015sja,Calibbi:2015kma,Barbieri:2016las,DiLuzio:2017vat,Calibbi:2017qbu} or by combining two scalar leptoquarks~\cite{Crivellin:2017zlb}. Neglecting small CKM factors, the assumption $C^{(1)}_{23}\approx C^{(3)}_{23}$ implies that contributions to $b\to c\tau^-\bar\nu_\tau$ and $b\to s\tau^+\tau^-$ are generated together in the combination
\begin{equation}
[\bar{c}_L\gamma_\mu b_L][\bar{\tau}_L\gamma^\mu\nu_\tau]+[\bar{s}_L\gamma_\mu b_L][\bar{\tau}_L\gamma^\mu\tau_L].
\end{equation}
This correlation means that effects in $b\to s\tau^+\tau^-$ are of the same order as the ones required to explain $R_{D^{(*)}}$, i.e., of the order of a tree-level SM process. We may neglect Cabibbo-suppressed contributions and assume that the NP contribution to $b\to c\tau^-\bar\nu_\tau$ is small compared
to the SM one, so that we keep only the SM contribution and the SM-NP interference terms in $b\to c\tau^-\bar\nu_\tau$ decay rates. We find the relation
\begin{equation}\label{eq:correlation}
C_{9(10)}^{\tau\tau}\,\approx\,C_{9(10)}^{\rm SM}-(+) \Delta\,,
\end{equation}
with
\begin{equation}\label{eq:YX}
  \Delta \,=\, \dfrac{2\pi}{\alpha}\frac{V_{cb}}{V_{tb}V_{ts}^*}\left( \sqrt {\frac{R_X}{R_X^{\rm SM}}}   - 1 \right) \,.
\end{equation}
In our framework, $\Delta$ is independent of the exclusive $b\to c\ell^-\bar\nu_\ell$ channel chosen $X$, see Eq.~(\ref{eq:ratiosR}).
Note that this prediction for the Wilson coefficients $C_{9}^{\tau\tau}$ and $C_{10}^{\tau\tau}$ is model independent, in the sense that the only ingredients in the derivation are the assumptions that NP only affects left-handed quarks and leptons and that it couples significantly to the second generation in such a way that experimental constraints can be avoided. 

We stress that the factor multiplying the bracket in Eq.~(\ref{eq:YX}) is very large (around 860). Using the current values for $R_{D^{(*)}}$, we obtain a positive (respectively negative) NP contribution to the Wilson coefficient $C_9^{\tau\tau}$ (respectively $C_{10}^{\tau\tau}$) parametrised by $\Delta=O(100)$ which overwhelms completely the SM contribution to these Wilson coefficients. Such large values of the Wilson coefficients are not in contradiction with the constraints obtained in Ref.~\cite{Bobeth:2011st} (when comparing with the results of this reference, one must be aware of the different normalisations of the operators in the effective Hamiltonian).

In view of these huge coefficients, we provide predictions for the relevant decay rates assuming that they are completely dominated by the NP contribution $\Delta$, and thus neglecting both short- and long-distance SM contributions. We obtain the branching ratios of the various 
$b\to s\tau^+\tau^-$ channels
%\begin{widetext}
\begin{eqnarray}\label{eq:bstautauNP}
{\rm{Br}}\left( B_s \to \tau^+ \tau^- \right) &=& 
\left(\frac{\Delta}{C^{\rm SM}_{10}}\right)^2\!\!\! {\rm{Br}}\left( {B_s} \to \tau^+ \tau^-   \right)_{\rm SM}\;\;\;\;
\end{eqnarray}
\begin{eqnarray}
{\rm{Br}}\left( {{B} \to K \tau^+ \tau^-} \right) &=&(8.8\pm 0.8)\times 10^{-9} \Delta^2 \,,\\
{\rm{Br}}\left( {{B} \to K^* \tau^+ \tau^-} \right) &=&(10.1\pm0.8)\times 10^{-9} \Delta^2\,,\\
		{\rm{Br}}\left( {{B_s} \to \phi \tau^+ \tau^-} \right) &=& (9.1\pm0.5) \times 10^{-9}\Delta^2\,,
\end{eqnarray}
%\end{widetext}
where the last three branching ratios are considered over the whole kinematic range for the lepton pair invariant mass $q^2$ (i.e., from $4m_\tau^2$ up to the low-recoil endpoint). We neglect the contributions only due to the SM. In the above expressions, the uncertainties quoted come from hadronic contributions multiplied by the short-distance NP contribution $\Delta$. A naive estimate suggests that the contribution of the $\psi(2S)$ resonance to this branching ratio amounts to $2\times 10^{-6}$, which is negligible in the limit of very large NP contributions considered here.
We thus may calculate the branching ratios for the whole kinematically allowed $q^2$ region, from the vicinity of the $\psi(2S)$ resonance up to the low-recoil endpoint, assuming that the result is completely dominated by the NP contribution. 

Since we neglected all errors related to the SM contribution for the semileptonic processes, we should do the same for $B_s \to \tau^+ \tau^-$.
For ${\rm{Br}}\left( {B_s} \to \tau^+ \tau^-   \right)_{\rm SM}$  in 
 Eq.~(\ref{eq:bstautauNP}), we should only consider the uncertainties coming from the $B_s$ decay constant and decay width as well as the different scales used to compute the Wilson coefficients here and in Ref.~\cite{Bobeth:2013uxa}, leading to a relative uncertainty of 4.7\% (to be compared with the larger 6.4\% uncertainty in Eq.~(\ref{brbstautauSM}) that includes other sources of uncertainties irrelevant under our current assumptions).

In Fig.~\ref{fig:predsbstautau}, we indicate the corresponding predictions as a function of $R_X/R_X^{\rm SM}$ (assumed to be independent of the $b\to c\ell^-\bar\nu_\ell$ hadronic decay channel $X$ in our approach). 
We have also indicated the current experimental range for $R_X/R_X^{SM}$, obtained by performing the weighted average of $R_D$, $R_{D^*}$ and $R_{J/\psi}$ without taking into account correlations.
We see that the branching ratios for semileptonic decays can easily reach $3\times 10^{-4}$, whereas $B_s\to\tau^+\tau^-$ can be increased up to $10^{-3}$.

Up to now, we have discussed the correlation between NP in $b\to c\tau\bar\nu_\tau$ and $b\to s\tau^+\tau^-$ under a limited set of assumptions that are fairly model independent. A comment is in order concerning the implications of these assumptions for $b\to s\mu^+\mu^-$. If we assume that the same mechanism is at work for muons and taus, we obtain also a correlation between $b\to s\mu^+\mu^-$ and $b\to c\mu^-\bar\nu_\mu$: the $O(25\%)$ shift needed in $C_{9}^{\mu\mu}$ and $C_{10}^{\mu\mu}$ to describe $b\to s\mu^+\mu^-$ data~\cite{Capdevila:2017bsm} translates into a very small positive $\Delta$ and a decrease of $b\to c\mu^-\bar\nu_\mu$ decay rates compared to the SM by a negligible amount of only a few per mille, so that there would be no measurable differences between electron and muon semileptonic decays. 
\bigskip

\section{Conclusions}
\label{Sec:conclusion}

In this article, we have studied the possibility of finding NP in $b\to s\tau^+\tau^-$ processes motivated by the converging experimental evidence for LFU violation in $b$-decays for both $b\to s$ and $b\to c$ transitions. 
We have updated the SM predictions for
${{B} \to K \tau^+ \tau^-}$, ${{B} \to K^* \tau^+ \tau^-}$ and ${{B_s} \to \phi \tau^+ \tau^-}$
and calculated the expression of these branching ratios
in terms of NP contributions to the $b\to s\tau^+\tau^-$  Wilson coefficients $C_{9}^{\tau\tau}$, $C_{10}^{\tau\tau}$, $C_{9'}^{\tau\tau}$ and $C_{10'}^{\tau\tau}$.

We have also analysed the correlation between NP contributions to $b\to s\tau^+\tau^-$ and $b\to c\tau^-\bar\nu_\tau$ under  general assumptions  in agreement with experimental indications: the deviations in $b\to c\tau^-\bar\nu_\tau$ decays come from a NP contribution to the left-handed four-fermion vector operator, this NP contribution is due to physics coming from a scale significantly larger than the electroweak scale, and the resulting contribution to $b\to s\nu_\tau\bar\nu_\tau$ is suppressed. 

Under these assumptions, an explanation of $R_{D^{(*)}}$ requires an enhancement of all $b\to s\tau^+\tau^-$ processes by approximately three orders of magnitude compared to the SM. In this case, the predictions for the branching ratios are completely dominated by NP contributions
and can be expressed simply as
\begin{eqnarray}
{{{\rm{Br}}\left( {{B_s} \to \tau^+ \tau^-}	\right)}} &\approx& { 0.031}\left( \sqrt{\frac {R_X}{R_X^{\rm SM}}}-1\right)^2\,,\end{eqnarray}

\begin{eqnarray}
		{\rm{Br}}\left( {{B} \to K \tau^+ \tau^-} \right)&\approx&0.007\left( \sqrt{\frac {R_X}{R_X^{\rm SM}}}-1\right)^2\,,\end{eqnarray}
		
		\begin{eqnarray}
		{\rm{Br}}\left( {{B} \to K^* \tau^+ \tau^-} \right)&\approx&0.008\left( \sqrt{\frac {R_X}{R_X^{\rm SM}}}-1\right)^2\,,		\end{eqnarray}
		
		\begin{eqnarray}
		{\rm{Br}}\left( {{B_s} \to \phi \tau^+ \tau^-} \right)&\approx&0.007\left( \sqrt{\frac {R_X}{R_X^{\rm SM}}}-1\right)^2\,.
\end{eqnarray}
with uncertainties of a few percent,
when integrated over the whole kinematic region allowed for the dilepton invariant mass. The corresponding enhancement yields branching ratios between $10^{-4}$ and $10^{-3}$ for these modes, as illustrated in Fig.~\ref{fig:predsbstautau}.

Our study confirms the potential of $b\to s\tau^+\tau^-$ decays to look for NP in the context of the measurements searching for violation of LFU in semileptonic $b$-decays. It is thus highly desirable to look for these decays in the current and forthcoming experiments studying $b$-decays such as LHCb and Belle II, which will provide complementary analyses of these decays with the exciting opportunity to discover NP in these transitions.

\acknowledgments{
This work received financial support from the grant FPA2014-61478-EXP [JM, SDG, BC]   and from Centro de Excelencia Severo Ochoa SEV-2012-0234 [BC]; from the EU Horizon 2020 program from the grants No 690575, No 674896 and No. 692194 [SDG]. The work of A.C. is supported by an Ambizione Grant of the Swiss National Science Foundation (PZ00P2\_154834).}

%\clearpage
\bibliography{BIB}

\end{document}